\documentstyle[11pt,amssymb,epsf]{article}
\newcommand{\be}{\begin{equation}}
\newcommand{\ee}{\end{equation}}
\newcommand{\ba}{\begin{array}}
\newcommand{\ea}{\end{array}}
\newcommand{\bea}{\begin{eqnarray}}
\newcommand{\eea}{\end{eqnarray}}

\renewcommand{\l}{\newline\null}

\newcommand{\rar}{\rightarrow}
\newcommand{\p}{\partial}
\newcommand{\ol}{\overline}

\newcommand{\la}{\langle}
\newcommand{\ra}{\rangle}
\def\figskip{\vskip .5cm plus 3mm minus 2mm}
\def\hbar{h\!\!\!/}

\begin{document}
\begin{titlepage}
%
April 1995\hfill PAR-LPTHE 95/08
\begin{flushright} hep-ph/9504201 \end{flushright}
\vskip 4cm
\begin{center}
{\bf ABOUT COMPOSITE SCALAR REPRESENTATIONS OF THE ELECTROWEAK
SYMMETRY GROUP OF THE STANDARD MODEL.}
\end{center}
\vskip .5cm
\centerline{B. Machet,
     \footnote[1]{Member of `Centre National de la Recherche Scientifique'.}
     \footnote[2]{E-mail: machet@lpthe.jussieu.fr.}
     }
\medskip
\centerline{{\em Laboratoire de Physique Th\'eorique et Hautes Energies,}
     \footnote[3]{LPTHE tour 16\,/\,1$^{er}\!$\'etage,
          Universit\'e P. et M. Curie, BP 126, 4 place Jussieu,
          F 75252 PARIS CEDEX 05 (France).}
}
\centerline{\em Universit\'es Pierre et Marie Curie (Paris 6) et Denis
Diderot (Paris 7);}\centerline{\em  Unit\'e associ\'ee au CNRS D0 280.}
\vskip 1.5cm
{\bf Abstract:}  the scalar composite representations of the electroweak
symmetry group of the Standard Model are exhibited in the case of two
generations of quarks.  The link between `strong' and `electroweak'
eigenstates is investigated, showing that the quark content commonly
attributed to pseudoscalar mesons needs to be modified.
After doing so, the mechanism suppressing the $K^+ \rar \pi^+\pi^0$ decays
with respect to $K_S \rar \pi^+\pi^-$ is unraveled, without reference to any
model of quark spectator or `factorization' hypothesis. The same
mechanism is shown to suppress $K_L \rar \pi\pi$ decays. Leptonic and
semi-leptonic decays are also studied. The mass relations between
`electroweak' eigenstates, verified at better than one percent,
$m_{\pi^3}/m_{\pi^\pm}=\cos\theta_c=m_{\chi(1910)}/m_{D_s^\pm}$ are obtained.
\smallskip

{\bf PACS:} 12.15.-y, 11.30.Rd, 12.15.Ji, 12.60.-i, 12.60.Fr, 12.60.Rc,
13.20.-v, 13.25.-k
\vfill
\null\hfil\epsffile{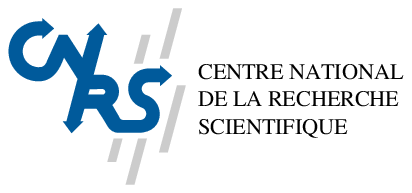}
\end{titlepage}
\section{Introduction.}

Accounting for the Cabibbo rotation \cite{Cabibbo} is customary at the level
of elementary
fields (quarks) when they undergo weak interactions; however, the
determination of the electroweak eigenstates for the observed mesons,
considered as composite particles, has, surprisingly, not been carefully
looked at, with the only exception, and still, as we shall see, incomplete, of
the neutral $K$ mesons. Yet, nearly all observed interactions of
pseudoscalar mesons are electroweak, which means that we have never observed
but electroweak eigenstates, and not the bound states associated with the
chiral group of symmetry of strong interactions \cite{Gell-Mann}.
The latter only play a role
in the creation of quark-antiquark pairs, which then evolve in time, and
 decay, according with the dynamics of electroweak interactions
\cite{GlashowSalamWeinberg}.
The theoretical investigations concerning mesonic decays could only
depart from the level of their basic constituents by the techniques of
Current Algebra \cite{CurrentAlgebra}, which are of reduced utility in
particular when non-leptonic decays are
concerned. The most striking evidence of the inadequacy of the usual line of
approach is probably the experimental factor 20 suppressing the amplitude
for the decay
$K^+ \rar \pi^+\pi^0$ with respect to $K_S \rar \pi^+\pi^-$ \cite{Kpipi},
totally unexpected in a description at the quark level, where the breaking of
the isospin symmetry between the $u$ and $d$ quarks seems the only factor which
can distinguish between the two reactions. Despite continuous efforts
\cite{Kpipi}, the
introduction of strong interactions, called for assistance in the form of
Quantum Chromodynamics \cite{QCD}, could never give a very clear explanation
for this enigmatic factor.

{\em Remark:} it is likely, as can be seen from the ``Table of
Particle Properties'' \cite{ParticleData},
that scalar mesons can decay by strong interactions,
and that the nature of the observed states be, in this case, better
described in terms of `strong' eigenstates. This is why we shall rather
focus on pseudoscalar states in the present work.

The first step of the approach to those reactions that is proposed here is
to identify the states which interact, and which are experimentally detected:
the 32 mesons (16 pseudoscalars and 16 scalars) that
can be built with 4 quarks fall into 8 representations of dimension 4 of the
$SU(2)_L \times U(1)$ symmetry group of the Standard Model;
four of them contain 1 scalar
and 3 pseudoscalars, and the other four contain 1 pseudoscalar and 3
scalars. New results are in particular obtained for what are known as `long'
and `short' lived kaons.\l
All entries in those composite representations now become functions of the
mixing angle (here the Cabibbo angle since we restrict the present study to
the case of two generations of quarks); this reflects the dependence on
this angle of the action of the electroweak group itself on the fundamental
fermions, in the natural basis of their `strong' eigenstates, {\em i.e.} the
quarks: in this basis, the generators of $SU(2)_L \times U(1)$ are $N\times N$
matrices ($N$ is the number of quark flavours, $N=4$ in this work),
with a projector
$(1-\gamma_5)/2$ for the `left' group $SU(2)$. In this way, the electroweak
group appears as a subgroup of the chiral $U(N)_L \times U(N)_R$ group of
the strong interactions.

Having identified the interacting states, in particular through their
leptonic and semi-leptonic decays, one then tackles the problem of the
non-leptonic decays of kaons, which do not take place through weak interactions
only, but also require the transformation, by strong interactions, of
a scalar state into two pseudoscalars, or that of a pseudoscalar into one
scalar and one pseudoscalar. Indeed, the only property of strong
interactions that we shall postulate is that they occur by the creation of a
quark-antiquark pair, diagonal in flavour. The details of the dynamics we
consider to be unknown, and we shall never rely on any QCD-like assumption.
The net result is that, in the limit of exact $SU(4)$ symmetry,
two types of diagrams contribute to $K^+ \rar \pi^+\pi^0$ decays, and
exactly cancel, while only one is present for $K_S \rar \pi^+\pi^-$;
the same mechanism yields the suppression of $K_L \rar \pi \pi$ decays.
Getting numerical values for the decay rates would requires the knowledge of
the strong vertex, which is not the case; if convergence is likely to require
that it behaves like an inverse power of the incoming momentum, one has always
to take that kind of argumentation {\em cum grano salis}. Anyhow, such as it
is,
the origin of the suppression is  unambiguous. It it based on a (gauge) theory
of scalar composite multiplets which includes the Standard Model, and is
free of any assumption like a `quark-spectator' model,
or/and of a factorization hypothesis.

Using a (gauge) electroweak theory for composite particles
would not be consistent without a justification to
eliminate the quark diagrams that have been up to now considered; this is
why we shall, at the end of the paper, briefly restate an argumentation
for quantizing spontaneously broken theories with non-independent degrees of
freedom \cite{Machet1,Machet2,BellonMachet} (mesons and
quarks), which shows that the quarks tend to decouple by getting an infinite
mass, their degrees of freedom being transmuted into those of the mesons.
Though the argument has only been worked out at leading order in $1/N$, the
results of the present work point in the same direction.

\section{The composite electroweak eigenstates of the standard model.}

\subsection{Preliminaries.}\label{sec:PRELIM}

For the sake of completeness and in order to make this work more easily
understandable in itself, we start by reproducing a few statements already
contained in ref.~\cite{Machet2}.

Let $\Psi$ be the fermionic 4-vector
\be
     \Psi = \left ( \ba{c}  u \\
                            c \\
                            d \\
                            s
           \ea \right ).
\label{eq:PSI}\ee
In this basis, the generators of the $SU(2)_L \times U(1)$ group, as
deduced from the Lagrangian of the Standard Model, act as follows:
\be
{\Bbb T}_L^+={\Bbb T}_L^1+i{\Bbb T}_L^2={{1-\gamma_5}\over 2}\ {\Bbb C}\ ,
\ {\Bbb T}_L^-={\Bbb T}_L^1-i{\Bbb T}_L^2={{1-\gamma_5}\over 2}\
{\Bbb C}^{\dag},
\ {\Bbb T}_L^3={1\over2}{{1-\gamma_5}\over2}\ {\Bbb N}\ ;
\ee
\quad - for $U(1)_L$:
\be
{\Bbb Y}_L={{1-\gamma_5}\over 2}\ {\Bbb Y}={1\over 6}
{{1-\gamma_5}\over 2}\ {\Bbb I};
\ee
\quad - for $U(1)_R$:
\be
{\Bbb Y}_R={{1+\gamma_5}\over 2}\ {\Bbb Y}={{1+\gamma_5}\over 2}\ {\Bbb Q}\:,
\ee
where
\be
{\Bbb C}=\left(\begin{array}{ccc}
                        0 & \vline & {\bf C}\\
                        \hline
                        0 & \vline & 0           \end{array}\right);
{\Bbb N}=\left(\begin{array}{ccc}
                        1 & \vline & 0\\
                        \hline
                        0 & \vline & {-1}        \end{array}\right);
{\Bbb Q}=\left(\begin{array}{ccc}
                        {2\over 3} & \vline & 0\\
                        \hline
                        0 & \vline & -{1\over 3}  \end{array}\right).
\ee
${\bf C}$ is the customary $2\times 2$ Cabibbo mixing matrix
\be {\bf C}= \left ( \ba{rr} \cos\theta_c & \sin\theta_c \\
                            -\sin\theta_c & \cos\theta_c    \ea \right ).
\label{eq:CABIBBO}\ee
The Gell-Mann-Nishijima relation writes:
\be
{\Bbb Y}={\Bbb Y}_R +{\Bbb Y}_L= {\Bbb Q}-{\Bbb T}^3_L={\Bbb Q}_R + {\Bbb Y}_L.
\label{eq:GMN}
\ee
The action of any among such generators, $\Bbb T$ or ${\Bbb T}\gamma_5$, on a
composite scalar $\ol\Psi A \Psi$ or pseudoscalar $\ol\Psi \gamma_5 A \Psi$,
where $A$ is a $4\times 4$ matrix, is
deduced by acting with the group on the fermions. Writing
\be
 e^{i\vec\alpha.\vec{\Bbb T}}.{\ol\Psi} A \Psi = {\ol\Psi} A \Psi +
                        i(\vec\alpha.\vec{\Bbb T}).{\ol\Psi} A \Psi + \ldots,
\ee
we get
\be
{\Bbb T}. {\ol\Psi} A \Psi ={\ol\Psi} [A,{\Bbb T}] \Psi,
\label{eq:GA1}\ee
and, similarly,
\bea
& &{\Bbb T}. {\ol\Psi} A \gamma_5 \Psi =
                            {\ol\Psi} [A,{\Bbb T}]\gamma_5 \Psi;\nonumber\\
& &{\Bbb T}\gamma_5 . {\ol\Psi} A \Psi =
                            {\ol \Psi} \{A,{\Bbb T}\}\gamma_5 \Psi;\nonumber\\
& &{\Bbb T}\gamma_5 . {\ol\Psi} A \gamma_5 \Psi=
                                   {\ol \Psi} \{A,{\Bbb T}\} \Psi,
\label{eq:GA2}\eea
where $[\ ,\ ]$ stands for `commutator' and $\{\ ,\ \}$ stands for
`anticommutator'. As both are involved when acting with the electroweak
group, any of its composite representation is associated with an {\em algebra
of matrices} (it closes for commutation and anticommutation and thus for simple
matrix multiplication). The simplest is formed by the three $SU(2)$ generators
${\Bbb T}^3, {\Bbb T}^+, {\Bbb T}^-$ and the unit matrix $\Bbb I$, and is at
the origin of the first among the eight representations that we
display below.

In the following, the symmetry is supposed to be spontaneously broken by
\be
\la H \ra = {v\over\sqrt{2}},
\ee
equivalent to
\be
\la \ol\Psi \Psi \ra = N\mu^3
\ee
by the relation
\be
H = {v\over\sqrt{2}N\mu^3} \ol\Psi \Psi.
\ee
{\em Remark:} in the whole paper, we suppose that the only fields with
non-vanishing vacuum expectation values are the diagonal scalar diquark
operators. This could have to be modified in a parity-violating theory like
that under scrutiny.

All representations below are contained in those displayed in
ref.~\cite{Machet2}, and can be easily deduced from them by simple
algebra. The expressions that I propose here are somewhat easier to
manipulate than the complex representations of the previous work.

The notations $\sin\theta_c, \ \cos\theta_c$ are hereafter replaced
with $s_\theta$ and $c_\theta$ respectively.

\subsection{The composite scalar representations of
{\boldmath $SU(2)_L\times U(1)$}.}\label{sec:REPS}

They write:

\vbox{
\bea
\Phi &=& \left( H,\phi^3,\phi^+,\phi^- \right)
        = {v\over N\mu^3}\ol\Psi\left(
            {1\over\sqrt{2}}{\Bbb I}, {i\over\sqrt{2}}\gamma_5{\Bbb N},
             i\gamma_5{\Bbb C},i\gamma_5{\Bbb C}^{\dag}
\right)\Psi = \nonumber\\
& &\hskip -2cm {v\over N\mu^3}\ol\Psi\left[
{1\over\sqrt{2}} \left(\ba{ccccc} 1 &   &\vline &    &    \nonumber\\
                                    & 1 &\vline &    &    \nonumber\\
                                    \hline
                                    &   &\vline &  1 &    \nonumber\\
                                    &   &\vline &    &  1 \ea \right),
{i\over\sqrt{2}}\gamma_5 \left(\ba{ccccc} 1 &   &\vline &   &   \nonumber\\
                                            & 1 &\vline &   &   \nonumber\\
                                           \hline
                                            &   &\vline &-1 &   \nonumber\\
                                            &   &\vline &   & -1   \ea \right),
\right .\nonumber\\
& & \hskip 6cm \left .
i\gamma_5 \left(\ba{ccccc}   &  &\vline & c_\theta &  s_\theta \nonumber\\
                             &  &\vline &-s_\theta &  c_\theta \nonumber\\
                            \hline
                             &  &\vline &   &     \nonumber\\
                             &  &\vline &   &  \ea \right),
i\gamma_5 \left(\ba{ccccc}   &   &\vline &   &   \nonumber\\
                             &   &\vline &   &   \nonumber\\
                             \hline
                             c_\theta &-s_\theta &\vline &   &   \nonumber\\
                             s_\theta & c_\theta &\vline &   &   \ea \right)
\right] \Psi; \nonumber\\
& &
\label{eq:PHI}
\eea
}
\vbox{
\bea
\Sigma &=& \left(\Sigma^0,\Sigma^3,\Sigma^+,\Sigma^- \right) = \nonumber\\
& &\hskip -2cm {v\over N\mu^3}\ol\Psi\left[
{1\over\sqrt{2}} \left(\ba{ccccc}
     1 &   &\vline &    &    \nonumber\\
       & -1 &\vline &    &    \nonumber\\
     \hline
  &  & \vline & c_\theta^2 - s_\theta^2 & 2c_\theta s_\theta  \nonumber\\
  &  & \vline & 2c_\theta s_\theta & s_\theta^2 -c_\theta^2  \ea \right),
{i\over\sqrt{2}}\gamma_5 \left(\ba{ccccc}
             1 &    & \vline &   &   \nonumber\\
               & -1 & \vline &   &   \nonumber\\
             \hline
      &  & \vline & s_\theta^2 - c_\theta^2 & -2c_\theta s_\theta  \nonumber\\
      &  & \vline & -2c_\theta s_\theta & c_\theta^2 - s_\theta^2  \ea \right),
\right .\nonumber\\
& & \hskip 6cm \left .
i\gamma_5 \left(\ba{ccccc}   &  &\vline & c_\theta & s_\theta \nonumber\\
                             &  &\vline & s_\theta & -c_\theta \nonumber\\
                            \hline
                             &  &\vline &   &     \nonumber\\
                             &  &\vline &   &  \ea \right),
i\gamma_5 \left(\ba{ccccc}     &   &\vline &   &   \nonumber\\
                               &   &\vline &   &   \nonumber\\
                              \hline
                              c_\theta & s_\theta &\vline &   &   \nonumber\\
                              s_\theta & -c_\theta &\vline &   &   \ea \right)
\right] \Psi; \nonumber\\
& &
\label{eq:SIGMA}
\eea
}
\vbox{
\bea
\Xi &=& \left(\Xi^0,\Xi^3,\Xi^+,\Xi^- \right) = \nonumber\\
& &\hskip -2cm {v\over N\mu^3}\ol\Psi\left[
{i\over\sqrt{2}} \left(\ba{ccccc}           & 1 &\vline &    &    \nonumber\\
                                         -1 &   &\vline &    &    \nonumber\\
                                            \hline
                                            &   &\vline &    & 1  \nonumber\\
                                            &   &\vline & -1 &    \ea \right),
{1\over\sqrt{2}}\gamma_5 \left(\ba{ccccc}   & 1 &\vline &   &   \nonumber\\
                                         -1 &   &\vline &   &   \nonumber\\
                                           \hline
                                            &   &\vline &   & -1 \nonumber\\
                                            &   &\vline & 1 &    \ea \right),
\right .\nonumber\\
& & \hskip 6cm \left .
\gamma_5 \left(\ba{ccccc}   &  &\vline & -s_\theta & c_\theta \nonumber\\
                            &  &\vline & -c_\theta & -s_\theta \nonumber\\
                            \hline
                            &  &\vline &   &     \nonumber\\
                            &  &\vline &   &  \ea \right),
\gamma_5 \left(\ba{ccccc}     &   &\vline &   &   \nonumber\\
                              &   &\vline &   &   \nonumber\\
                            \hline
                           s_\theta & c_\theta &\vline &   &   \nonumber\\
                           -c_\theta & s_\theta &\vline &   &   \ea \right)
\right] \Psi; \nonumber\\
& &
\label{eq:XI}
\eea
}
\vbox{
\bea
\Omega &=& \left(\Omega^0,\Omega^3,\Omega^+,\Omega^- \right) = \nonumber\\
& &\hskip -2cm {v\over N\mu^3}\ol\Psi\left[
{1\over\sqrt{2}} \left(\ba{ccccc}
        & 1 &\vline &    &    \nonumber\\
      1 &  &\vline &    &    \nonumber\\
     \hline
     &  & \vline & -2c_\theta s_\theta & c_\theta^2 - s_\theta^2   \nonumber\\
     &  & \vline &  c_\theta^2 - s_\theta^2 & 2c_\theta s_\theta   \ea \right),
{i\over\sqrt{2}}\gamma_5 \left(\ba{ccccc}
         & 1 & \vline &   &   \nonumber\\
       1 &  & \vline &   &   \nonumber\\
       \hline
     &  & \vline & 2c_\theta s_\theta & s_\theta^2 - c_\theta^2  \nonumber\\
     &  & \vline & s_\theta^2 - c_\theta^2 & -2c_\theta s_\theta \ea \right),
\right .\nonumber\\
& & \hskip 6cm \left .
i\gamma_5 \left(\ba{ccccc}   &  & \vline & -s_\theta & c_\theta \nonumber\\
                             &  & \vline & c_\theta &  s_\theta \nonumber\\
                            \hline
                            &  &\vline &   &     \nonumber\\
                            &  &\vline &   &  \ea \right),
i\gamma_5 \left(\ba{ccccc}     &   &\vline &   &   \nonumber\\
                               &   &\vline &   &   \nonumber\\
                              \hline
                             -s_\theta & c_\theta &\vline &   &   \nonumber\\
                              c_\theta & s_\theta &\vline &   &   \ea \right)
\right] \Psi. \nonumber\\
& &
\label{eq:OMEGA}
\eea
}
The other four representations are deduced from the four above by a
$\gamma_5$ transformation, and contain one pseudoscalar and three scalars:
\bea
& & \tilde\Phi = \Phi\ (scalar \longleftrightarrow pseudoscalar), \nonumber\\
& & \tilde\Sigma = \Sigma\ (scalar\longleftrightarrow pseudoscalar),\nonumber\\
& & \tilde\Xi = \Xi\ (scalar \longleftrightarrow pseudoscalar), \nonumber\\
& & \tilde\Omega = \Omega \ (scalar \longleftrightarrow pseudoscalar).
\eea
{\em Remark:} All `untilded' representations, as have been written, are
hermitian  (and all `tilded' representations antihermitian),
which explains  the difference in the $i$ factors that can be noticed between
$\Xi$ and $\Phi, \Sigma, \Omega$.

On any representation $(\phi^0, \vec \phi)$, the group acts as follows:
\bea
{\Bbb T}^i_L. \phi_j & = &
-{i\over 2}(\varepsilon_{ijk}\phi^k + \delta_{ij}\phi^0), \nonumber\\
{\Bbb T}^i_L. H & = & {i\over 2}\phi_i\;.
\label{eq:GA3}\eea
We have as usual
\be
\phi^+={\phi_1+i\phi_2\over\sqrt{2}},\
\phi^-={\phi_1-i\phi_2\over\sqrt{2}}.
\ee
Any linear combination of the above representations is also a suitable
representation. This
leaves a certain freedom in the identification of the eigenstates with
observed particles. The resulting ambiguity can only be taken care of by
physical arguments.

To every representation above is associated a quadratic invariant (the
sum of the square of its four entries), and there are consequently {\em
a priori} eight independent electroweak mass scales; we shall see later that
they are finally reduced to six.

We also introduce the eight corresponding $SU(2)_L \times U(1)$ gauge-invariant
kinetic terms, all normalized to $1$; the one attached to $\Phi$ is
that for the usual scalar multiplet in the Glashow-Weinberg-Salam model.

{\em Remarks:}\l
- the sum of the kinetic terms for the electroweak eigenstates, all normalized
to $1$, is identical to that for the `strong' eigenstates $\bar u u,
\bar u \gamma_5 u, \bar u d, \bar u \gamma_5 d \ldots$ (up to the factor
$v/N\mu^3$), all likewise chosen with the same normalization; expressed for the
fields themselves, which transform like their covariant derivatives, it
reads:
\be
\vert\Phi\vert^2 + \vert\tilde\Phi\vert^2 +\vert\Sigma\vert^2 +
\vert\tilde\Sigma\vert^2 +\vert\Xi\vert^2 +\vert\tilde\Xi\vert^2 +
\vert\Omega\vert^2 +\vert\tilde\Omega\vert^2 =
2({v\over N\mu^3})^2
\sum_{q_i=u,c,d,s}\sum_{q_j=u,c,d,s}
\left( (\bar q_i q_j)^2 - (\bar q_i\gamma_5 q_j)^2 \right);
\label{eq:KINET}\ee
- the usual isospin group of strong interactions has generators
\be
I^+ = \left( \ba{ccccc}   &  & \vline & 1 &    \nonumber\\
                          &  & \vline &   & 0  \nonumber\\
                          \hline
                          &  & \vline &  &   \nonumber\\
                          &  & \vline &  &    \ea \right),\quad
I^- = \left( \ba{ccccc}   &  & \vline &  &    \nonumber\\
                          &  & \vline &  &    \nonumber\\
                          \hline
                        1 &  & \vline &  &   \nonumber\\
                          & 0 & \vline &  &    \ea \right),\quad
I^3 = {1\over 2} \left( \ba{ccccc}  1 &  & \vline &  &    \nonumber\\
                                      & 0 & \vline &  &    \nonumber\\
                                     \hline
                                      &  & \vline & -1 &   \nonumber\\
                                      &  & \vline &  & 0  \ea \right);
\label{eq:ISOSPIN}\ee
It can be generalized here to its `Cabibbo-rotated' version;\l
- the arrangement of scalars and pseudoscalars inside the representations,
always of the type $(1,3)$, sheds some light on the importance of the group
$SU(2)_{diagonal}$: inside each representation, the three scalars or the three
pseudoscalars are triplets of this group, while the remaining entry is a
singlet. $SU(2)_{diagonal}$ can also be considered as a generalization,
in the case of four quarks, of the isospin group of symmetry for the case of
two quarks;\l
- the eight composite representations can be deduced from one another by the
action of the $U(2)_L \times U(2)_R$ group with $U(2)$ generators $({\Bbb
I},\vec\tau)$:
\bea
& &\hskip -2cm
          \tau_1 = \left( \ba{ccccc}   & 1 &\vline &    &    \nonumber\\
                                 -1 &   &\vline &    &    \nonumber\\
                                    \hline
                                    &   &\vline &    & 1  \nonumber\\
                                    &   &\vline & -1 &    \ea \right),
          \tau_2 = \left(\ba{ccccc}
       & 1 &\vline &    &    \nonumber\\
      1 &  &\vline &    &    \nonumber\\
      \hline
        &   &\vline & -2c_\theta s_\theta & c_\theta^2 -s_\theta^2  \nonumber\\
        &   &\vline & c_\theta^2 -s_\theta^2 & 2c_\theta s_\theta  \ea \right),
          \tau_3 = \left(\ba{ccccc}
      1 &   &\vline &    &    \nonumber\\
       & -1 &\vline &    &    \nonumber\\
      \hline
       &   &\vline & c_\theta^2 -s_\theta^2 & 2c_\theta s_\theta \nonumber\\
       &   &\vline & 2c_\theta s_\theta & s_\theta^2 -
                              c_\theta^2 \ea\right).\nonumber\\
& &
\eea
Phrased in another way, the 16 matrices of the `Cabibbo rotated chiral $U(4)$
algebra' can be obtained
by simple matrix multiplication of the 4 matrices $({\Bbb I},\vec{\Bbb T})$
spanning the `electroweak $U(2)$' with the 4 matrices $({\Bbb I}, \vec \tau)$.
We have the commutation relations $[\vec \tau,\vec T] =0$.

\section{Linking observed pseudoscalar mesons and electroweak eigenstates.}

All quark pairs are first created by strong interactions, before their
electroweak interactions are observed. The incoming `strong' eigenstate is a
linear combination of electroweak mass eigenstates, with coefficients
depending on the mixing angle. The probability of observing a state with
definite (electroweak) mass thus also depends on the mixing angle. The
interactions and decays usually concern a precise mass state, which has
eventually been collimated.

\subsection{First steps.}\label{sec:FIRSTSTEPS}

Let us first investigate the nature of the electroweak mass eigenstates
called kaons. Suppose that, in a conservative way, we keep attributing to
the $K^+$ meson the quark content symbolized by $\bar u \gamma_5 s$.
It decomposes into
\be
i\,\bar u \gamma_5 s = {1\over 2}{N\mu^3\over v}
\left( c_\theta (i\,\Xi^+ + \Omega^+) + s_\theta (\Phi^+ + \Sigma^+)\right).
\label{eq:US}\ee
The neutral pseudoscalar in the same representation is expected to be
degenerate in mass with $\bar u \gamma_5 s$ and must be a linear combination
of $K^0$ and $\ol K^0$:
\be
{1\over 2}{N\mu^3\over v}
\left( c_\theta (i\,\Xi^3 + \Omega^3) + s_\theta (\Phi^3 + \Sigma^3) \right) =
{i\over\sqrt{2}}\left( c_\theta(\bar u\gamma_5 c - \bar d\gamma_5 s) +
s_\theta(\bar u\gamma_5 u - \bar s\gamma_5 s)\right).
\label{eq:KBAD}\ee
By the same argumentation, would we attribute to the $D^+$ meson the quark
content
$\bar c\gamma_5 d$, its neutral partner in the same representation would be
\be
{i\over\sqrt{2}}\left( c_\theta(\bar c\gamma_5 u - \bar s \gamma_5 d) -
               s_\theta (\bar c \gamma_5 c - \bar d\gamma_5 d)\right).
\label{eq:DBAD}\ee
Comparing the two expressions (\ref{eq:KBAD}) and (\ref{eq:DBAD}),
and neglecting corrections in $s_\theta$, we
see that the neutral partner of the $D^+$ meson and that of the $K^+$ are
antiparticles, and so {\em must} have the same
mass. Thus, unless we accept that the (large) mass splitting between
neutral $K$ and $D$'s is due to the (small) components, proportional
to $s_\theta$, of the neutral partners
of $\bar u \gamma_5 s$ and $\bar c \gamma_5 d$ which are
not antiparticles of each other, the starting hypothesis is untenable;
this  leads us  to associate the kaon mass to the
$\Xi$ multiplet and the $D$ mass to the $\Omega$ multiplet. The
study of leptonic, semi-leptonic and non-leptonic decays of these
weak eigenstates will further assert the fact that, with two generations of
quarks, the quark content of the $K$ and $D$ mesons cannot be what has
been currently believed. Instead, for example, we shall rather take as a
first approximation
\bea
K^+ &\approx & {i\over 2a}\Xi^+ \approx {i\,v\over 2aN\mu^3}
        (\bar u \gamma_5 s -\bar c\gamma_5 d),\nonumber\\
D^+ &\approx & {1\over 2a}\Omega^+ \approx {i\,v\over 2aN\mu^3}
        (\bar u \gamma_5 s + \bar c\gamma_5 d),
\label{eq:KDPLUS}\eea
where $a$ is a normalization constant that will be determined in section
\ref{sec:LEPTONIC} to be $a = 2f/v$; $f$ is the leptonic
decay constant, supposed, for simplification, to be the same for all
pseudoscalars.
The neutral $K$ and $D$ mesons will be found in `tilded' or `untilded'
representations, essentially discriminated by the presence or absence of
semi-leptonic decays (see section \ref{sec:SEMILEPT} below).

The case of the $\pi$ and $D_s$ mesons is different in that there is not the
same contradiction as above in taking a composition as close as possible to
$\bar u \gamma_5 d$ for $\pi^+$, and to $\bar c \gamma_5 s$ for $D_s^+$.
This becomes even mandatory, as seen below, for explaining the corresponding
leptonic and semi-leptonic decays, and also the non-leptonic decays of the
kaons. An argumentation, anticipating on next subsections, is the following:\l
- $\pi$ cannot be associated with $\Phi$ alone  because no neutral pseudoscalar
meson made of $d$ and $s$, or $u$ and $c$ ($\bar d\gamma_5 s, \bar s\gamma_5
d, \bar c\gamma_5 u,\bar u \gamma_5 c$) has any component on $\Phi$, and
thus will never have an element of $\Phi$ in the product of its semi-leptonic
decays;\l
- it cannot either be associated with $\Sigma$ alone as
the action of the group on any entry of $\Sigma$ never gives a scalar with
non-vanishing vacuum expectation value (we suppose, for the sake of
simplicity, that $\la \bar u u \ra = \la \bar c c\ra = \la \bar d d \ra =
\la \bar s s\ra$); this forbids the pure
$\Sigma$ states to leptonically decay. So, the pions, which do appear in
products of semi-leptonic decays of neutral strongly created quark pairs
$\bar d\gamma_5 s, \bar s\gamma_5 d, \bar c\gamma_5 u,\bar u \gamma_5 c$,
and which do decay leptonically, have to be mixtures of $\Phi$ and $\Sigma$;\l
- however, pions and $D_s$ mesons cannot be either combinations of $\Phi$ and
$\Sigma$ only, since then only pions and $D_s$ mesons could decay
leptonically, and not kaons; this leads us to introduce an admixture of
$\Xi$ in the definition of pions in particular (see eq.~(\ref{eq:MESONS})
below);\l
- then the mechanism at the origin of the suppression of $K^+ \rar \pi^+
\pi^0$ and $K_L \rar \pi \pi$ with respect to $K_S \rar \pi \pi$  can be
understood in simple terms.

Another mixing matrix is thus seen to occur, that relates the observed physical
states to the composite representations displayed in eqs.~(\ref{eq:PHI}) to
(\ref{eq:OMEGA}). It should be unitary, keeping the kinetic terms diagonal,
such that the mass terms are those that explicitly break the
$U(4)_L\times U(4)_R$ chiral symmetry down to $SU(2)_L\times U(1)$,
at the mesonic level. We shall see in section \ref{sec:PROPOSED} that it is
again the Cabibbo matrix, which now also acts at the level of composite states.

Keeping the pions and $D_s$ mesons as close as possible to their `strong'
composition
is akin to saying that bound states occurring in the first generation
only $(u,d$) or in the second generation only $(s,c)$ stay nearly unchanged
(up to corrections in $s_\theta$) when turning on
weak interactions, while it causes a strong departure
from the Gell-Mann $SU(3)$ symmetry; this is not surprising as the electroweak
gauge group cannot be embedded in chiral $U(3)_L \times U(3)_R$ while the
weak $SU(2)$ can be split into two $SU(2)$'s attached, up to $s_\theta$
corrections, respectively to the first and to the second  generation; the
first of these two groups is the (rotated) strong isospin group.

An interesting point is that the three pseudoscalar components of
$\Phi$ are known to become the third polarizations of the massive $W$'s
\cite{AbersLee}, which are naturally considered as pure electroweak states.
In this case where we only consider two generations of fermions, we seem to
be led, instead, to picture the third components of the massive gauge fields
as  mixtures of the quarks they are coupled to, {\em i.e.} essentially
the lowest $(\pi)$ and highest mass $(D_s)$ pseudoscalars, plus a small
(proportional to $s_\theta$) admixture of other mesons.
This is untenable, since admitting that the $W$ propagator has also a pole,
for example, at the pion mass is equivalent to saying that
there exist $W$-mediated strong interactions {\em also between hadrons and
leptons}. The problem has to be solved in the case of three generations, by
the requirement that the three pseudoscalar companions of the Higgs bosons,
which are three among the eleven pseudoscalars involving the top quark,
are pure mass states identical with the third components of the
massive vector bosons. This means in particular that three among the
eleven `topped' mesons weight around $80\ GeV$ and have already been observed.
The eight remaining ones may indeed have a higher mass scale \cite{top}.

So, from simple arguments, we have come to the conclusion that the
usual quark content attributed to pseudoscalar mesons has to be modified when
one turns on electroweak interactions for more than one generation of
quarks.

\subsection{Leptonic decays of pseudoscalar mesons.}\label{sec:LEPTONIC}

The leptonic decays of pseudoscalar mesons arise from the  couplings
$\propto g\la H \ra \vec W_\mu \p^\mu \vec\Phi$  occurring in the kinetic term
for $\Phi$,  with $\la H \ra = v/\sqrt{2}$; the outgoing leptons originate
from the decay of the $W$ gauge boson. All electroweak pseudoscalar
mass eigenstates making up $\Phi^3$ and $\Phi^\pm$ can thus decay into leptons;
this includes here,
in the charged sector, pions, kaons and $D_s$; the $D^\pm$ do not appear as
components of $\Phi^\pm$ (see eq.~(\ref{eq:MESONS}) below) and, indeed,
one does observe experimentally that
their leptonic decays are extremely small, smaller that those of the
$D_s^\pm$ (improving this picture to explain the existing small leptonic
decays of the $D$ mesons could be done for example by breaking the
postulated $SU(4)$ symmetry for the quark condensates, since the scalar
partner of the charged $D$'s involves the quantity $2\,c_\theta s_\theta\,
(\la\bar s s\ra - \la\bar d d\ra)$). Scalars cannot leptonically decay as
soon as no electroweak transformation can transmute them into a field with
non-vanishing vacuum expectation value, which is the case here as  we suppose
that only the diagonal scalar diquark operators condense in the vacuum, and are
$SU(4)$ symmetric.

Let for example a $(\bar u, s)$ quark pair be created by strong interactions
in a pseudoscalar state $\bar u \gamma_5 s$. Using the fact that the kinetic
terms, according to eq.~(\ref{eq:KINET}), can be identically expressed
in the `strong' or
in the `weak' basis, we directly study here the kinetic term for
$(iv/N\mu^3)\bar u \gamma_5 s$. Incorporating the factor $2$ which appears in
the r.h.s of eq.~(\ref{eq:KINET}), we get the coupling
\be
s_\theta {g\over 2} ({v\over  N\mu^3})^2\p^\mu (\bar u \gamma_5 s) W_\mu^+
(\la \bar u u + \bar s s\ra + \cdots).
\ee
Supposing $\la \bar u u\ra = \la \bar c c\ra = \la \bar d d\ra = \la \bar s
s\ra = N\mu^3 /4 = \mu^3$, it rewrites
\be
s_\theta{gv^2\over 4N\mu^3}\p^\mu (\bar u \gamma_5 s) W_\mu^+ + \cdots.
\ee
In the low energy regime, we can take for the $W$ propagator
$g_{\mu\nu}/M_W^2$, and, using $M_W^2 = g^2 v^2/8$, the diagram of fig.~1
finally yields the coupling
\be
s_\theta{2\over v}\p^\mu ({v\over N\mu^3}\bar u \gamma_5 s) L^-_\mu +\cdots,
\label{eq:CPL}\ee
where $L_\mu$ is the leptonic current, for example here
\be
L^-_\mu = \bar \mu \gamma_\mu {1-\gamma_5 \over 2} \nu_\mu.
\ee
\vbox{
\figskip
\hskip 4cm\epsffile{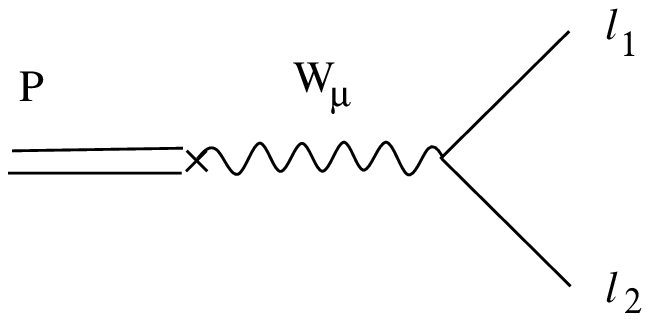}
\figskip
{\centerline{\em Fig.~1: leptonic decay of a pseudoscalar meson.}}
\figskip}
We now, as in ref.~\cite{Machet2}  rescale the fields by
\bea
\Psi &=& a\;\Psi', \nonumber\\
\Psi_{\ell}  &=& a\;\Psi'_{\ell}, \nonumber\\
R &=& a\; R' \ \mbox{for}\ R\in
\{\Phi,\Xi,\Sigma,\Omega,\tilde\Phi,\tilde\Xi,\tilde\Sigma,
                        \tilde\Omega\},\nonumber\\
\sigma_\mu &=& a\;\sigma'_\mu \ \mbox{for\ all\ gauge\ fields}\ \sigma_\mu,
\label{eq:SCALE}\eea
($\Psi_\ell$ is the leptonic equivalent of $\Psi$, eq.~(\ref{eq:PSI})),
the gauge coupling constants $g\cdots$ by
\be
g = g'/a \cdots,
\ee
and consider the classical Lagrangian ${\cal L}/a^2$, such that the kinetic
terms for the rescaled fields stay normalized to unity.
Notice that the gauge couplings
\be
g\,\phi_1 \p_\mu \phi_2 \sigma^\mu \ \mbox{or}
\ g\,\bar\Psi \gamma_\mu \Psi \sigma^\mu
\ee
keep the same form in the rescaled Lagrangian
\be
g'\,\phi'_1 \p_\mu \phi'_2 \sigma'^\mu \ \mbox{and}
\ g'\,\bar\Psi' \gamma_\mu \Psi' \sigma'^\mu
\label{eq:GAUGECPL}\ee
and, for low energy applications, that
\be
g^2 \la H\ra ^2 = g'^2 \la H'\ra^2.
\label{eq:GH}\ee
The physical fields and couplings are now considered to be the rescaled
(primed) objects.

Writing (see also eqs.~(\ref{eq:MESONS}) below)
\be
{iv\over N\mu^3}\bar u \gamma_5 s = a(K^+ + D^+),
\label{eq:NORME}\ee
the coupling (\ref{eq:CPL}) becomes, in the rescaled Lagrangian
\be
s_\theta{2 a\over v}\p^\mu (K^+ + D^+) L'^-_\mu +\cdots.
\label{eq:CPL2}\ee
In the same way, the kinetic term for $(iv/N\mu^3)\;\bar c\gamma_5 d$ yields
the coupling
\be
s_\theta{2 a\over v}\p^\mu (-K^+ + D^+) L'^-_\mu +\cdots.
\label{eq:CPL3}\ee
We consider, for the sake of simplicity, all leptonic decay constants to
be identical.

Summing the two contributions (\ref{eq:CPL2}) and (\ref{eq:CPL3})
we recover the same coupling
\be
s_\theta{8f\over  v^2} K^+ \p^\mu L'^-_\mu
\label{eq:PCAC1}\ee
as would be given by the usual PCAC computation,  if we take
\be
a = 2\;{f\over v} = {\sqrt{2} f \over v/\sqrt{2}}.
\ee
Indeed, eq.~(\ref{eq:PCAC1}) also writes in momentum space
\be
s_\theta\,g^2\,f\,(ip_\mu) {g^{\mu\nu}\over M_W^2} \bar\mu' \gamma_\mu
{1-\gamma_5\over 2} \nu_\mu',
\label{eq:PCAC2}\ee
where $p_\mu$ is the kaon momentum, yielding exactly the correct numerical
value $s_\theta f\,(ip_\mu)\,G_F$ for the leptonic coupling of $K$ to the
(physical)
`primed' leptons. $G_F$ is the Fermi constant $G_F = g^2/M_W^2= 8/v^2$.
The introduction of another scale of interactions \cite{SusskindWeinberg} is
not required by leptonic decays.

Note that putting together eqs.~(\ref{eq:NORME}) and (\ref{eq:SCALE})
we can express the mesons fields in terms of the `primed' (physical) quark
fields (this is again an approximation at ${\cal O}(s_\theta)$; see
eqs.(\ref{eq:MESONS}) below for the exact form proposed):
\bea
K^+  &\approx &
        i\,{f\over N\mu^3} (\bar u' \gamma_5 s' - \bar c' \gamma_5 d')
     = i\,{v'\over 2N\mu^{'3}} (\bar u' \gamma_5 s' - \bar c' \gamma_5 d')
     \approx {i\over 2} \Xi'^+,\nonumber\\
D^+  &\approx &
        i\,{f\over N\mu^3} (\bar u' \gamma_5 s' + \bar c' \gamma_5 d')
      = i\,{v'\over 2N\mu^{'3}} (\bar u' \gamma_5 s' + \bar c' \gamma_5 d')
      \approx {1\over 2} \Omega'^+,
\label{eq:KD}\eea
where
\be
v = a v' ,\  N\mu^3 = \la\ol\Psi\Psi\ra = a^2 N\mu^{'3} =
         a^2 \la \ol\Psi'\Psi'\ra.
\ee
Those expressions have to be compared with the customary `PCAC' relations
linking the $K$ and $D$ mesons
interpolating fields with the divergences of the corresponding axial currents
\bea
K^+ &=& {i(m_u + m_s)\over f_K M_K^2}\; \bar u \gamma_5 s,\nonumber\\
D^+ &=& {i(m_c + m_d)\over f_D M_D^2}\; \bar c \gamma_5 d,
\label{eq:PCAC3}\eea
$m_u, m_s, m_d$ and $m_c$ being the quark `masses' in the QCD Lagrangian.
Up to non-important numerical factors, we notice from eqs.~(\ref{eq:KD}) and
(\ref{eq:PCAC3}) that
$f/N\mu^3$ plays the same role, in the `rescaled' theory, as used
to do $m/f M^2$ for the original quark fields. This is
reminiscent of the Gell-Mann-Oakes-Renner \cite{GellMannOakesRenner} formula
$(m_u + m_s)N\mu^3 = f_K^2 M_K^2$.

\subsection{{\boldmath $\pi^0 \rar \gamma\gamma$} decays.}\label{sec:ANOM}

The $(\bar u \gamma_5 u - \bar d \gamma_5 d)$ bound state decomposes into
(we neglect here $s_\theta$ corrections to define the $\pi^0$ state)
\be
{iv\over N \mu^3}(\bar u \gamma_5 u - \bar d \gamma_5 d) = {1\over \sqrt{2}}
\left(\Phi^3 + \Sigma^3 + c_\theta s_\theta (i\,\tilde\Omega^0 - \Omega^3)
  + s_\theta^2 (i\,\tilde\Sigma^0 - \Sigma^3)\right).
\label{eq:UUDD}\ee
As shown in ref.~\cite{Machet1}, $\Phi^3$ decays into
two photons through the Lagrangian of constraint expressing its compositeness
in the Feynman path integral. Though the theory is anomaly-free,
the quarks becoming infinitely massive through the same constraint,
one rebuilds the same
amplitude for the decay $\pi^0 \rar \gamma\gamma$  as usually computed from
the anomalous divergence of the associated axial current.

All neutral states having a component on $\Phi^3$ will similarly decay into two
photons with an amplitude proportional to the Adler's anomaly \cite{Adler}.

{\em Remark:} the reader will easily make the transition from the abelian
exercise proposed in \cite{Machet1} and the realistic case studied here; he
may however  wonder why we only introduce constraints for the $\Phi$
multiplet,  expressing its compositeness, while one could expect, in this
non-abelian case, the equivalent for the other scalar multiplets;
this is because the constraints on $\Phi$ kill enough degrees of freedom to
yield a relevant theory. Its three pseudoscalar entries keep physical
because they are `eaten' by the gauge fields to become massive.
Putting constraints for the other representations
tend, at the opposite, to decouple their entries by giving them infinite
masses (unless they could be eaten by other gauge fields, meaning that one
could extend the gauge group to the full chiral group), which is unwanted.
This is why, in this framework, only the mesons which have a component
on $\Phi^3$ we expect to decay into two photons, with an amplitude
proportional to the anomaly.

\subsection{Semi-leptonic decays.}\label{sec:SEMILEPT}

Semi-leptonic decays (see fig.~2) are triggered by the same type of
couplings (occurring in the kinetic terms) as above, except that the action of
the gauge group on the initial pseudoscalar now yields another pseudoscalar
in the same representation (remember that the action of a {\em left}
generator on a pseudoscalar can yield a scalar as well as a pseudoscalar).
The $W$ gauge field still provides the two leptons. Remark that scalar
mesons can now decay semi-leptonically. Similarly, the pseudoscalar
`singlets' of `tilded' representations can only be turned into scalars by
the group, such that they will not have semi-leptonic decays into another
pseudoscalar. This is an important reason  why the $K_S$ meson has to mainly
belong to such a representation.
As above, we work with a precise example, now that of a strongly created
$\bar d \gamma_5 s$ pseudoscalar quark pair, which decomposes into
\be
\bar d \gamma_5 s = {N\mu^3\over v \sqrt{2}}\left(
     {1\over 2}(-i\,\tilde\Xi^0 - \Xi^3 + \tilde\Omega^0 +i\,\Omega^3) +
     c_\theta s_\theta(\tilde\Sigma^0 +i\,\Sigma^3) -
     s_\theta^2(\tilde\Omega^0 +i\,\Omega^3) \right).
\label{eq:DS}\ee
The weak interactions acting on the $\Sigma^3$ component can turn it into a
$\Sigma^+$ or $\Sigma^-$, essentially `made of' $\pi^\pm (\bar u\gamma_5
d,\bar d \gamma_5 u)$ and $D_s^\pm (\bar c\gamma_5 s, \bar s\gamma_5 c)$,
with a $s_\theta c_\theta$ factor. The strongly created $\bar d\gamma_5 s $
pair will thus semi-leptonically decay into $\pi^\pm \;+\;leptons$, with the
usual Cabibbo factor (up to $s_\theta^2$ corrections).
The other terms in the expansion (\ref{eq:DS}) can also give semi-leptonic
decays; the mass scales of the final states will determine if the decay
effectively takes place, depending on the energy available.
\vbox{
\figskip
\hskip 4.5cm\epsffile{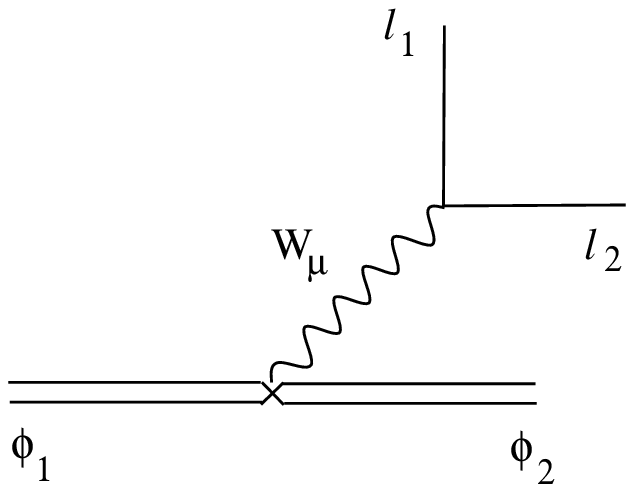}
\vskip 1mm
{\centerline{\em Fig.~2: semi-leptonic decay of a meson.}}
\figskip}
As already emphasized, this example is instructive in showing that the pions
cannot be identified with the three pseudoscalar entries of $\Phi$ since they
will never appear in the semi-leptonic decays of $\bar d \gamma_5 s$.
It should also be stressed that the semi-leptonic decays of neutral kaons
are experimentally
attributed to the long-lived one, while we rather attribute it to the pionic
component of $\bar d \gamma_5 s$ or $\bar s \gamma_5 d$, $\Sigma^3$, which
does not decay into two photons (remember that the neutral pion is a
mixture of $\Phi^3$ and $\Sigma^3$ plus small corrections, and only the
first component decays into two photons), and thus which is expected to have a
`lifetime' close to that of the charged pions, of the same order of
magnitude as that of the long-lived neutral kaon; this similarity between these
lifetimes, while that of $K_S$ is much smaller, makes reasonable to think
that there might be a `confusion' regarding the origin of the outgoing
meson.

Also, from the decomposition (\ref{eq:DS}) above, a $\bar d \gamma_5 s $ pair
is seen
to `split' mainly into four different mass eigenstates $\Xi^3,
\tilde\Xi^0, \Omega^3, \tilde\Omega^0$; the first two we identify will
the `long' and `short' lived neutral kaons, nearly degenerate in mass with
$K^\pm \equiv \Xi^\pm$, and the last two with the two neutral $D$ mesons,
nearly degenerate in mass with the charged ones.

\subsection{More about rescaling the fields and normalizing the
amplitudes.}

The rescaling (\ref{eq:SCALE}) deserves more comments.
\footnote{The remark on page 25 of the first version of ref.~\cite{Machet2}
involves an incorrect statement (suppressed in the revised version submitted
for publication), and should be replaced by the present subsection.}

In the rescaled Lagrangian ${\cal L}/a^2$ considered as the `classical'
Lagrangian for the rescaled `primed' fields and couplings,
the former appear normalized to $one$ and the gauge couplings take
exactly the same form (see eq.~(\ref{eq:GAUGECPL})) as they had
in the Lagrangian $\cal L$ in terms of the original fields and couplings;
we have seen that this led us to recover the correct leptonic decay amplitude
for pseudoscalar mesons.

However, when dealing with quantum effects, {\em i.e.} loops,
as for the $\pi^0 \rar \gamma\gamma$ decay, or with a
different number of asymptotic states, like for semi-leptonic decays,
this simple but somewhat too brutal procedure leads to problems of
normalization; I propose  instead to start from the generating functional
\be
{\cal Z} = \int {\cal D}\phi\;e^{{1\over \hbar c}\int d^4x {{\cal L}(\phi,g)}}
    \equiv \int {\cal D}\phi'\;e^{{a^2\over \hbar c}
                         \int d^4x {1\over a^2}{\cal L}(a\phi'\!,\;g'/a)}.
\label{eq:Z}\ee
where $\phi$ stands for all fields and $g$ for all gauge couplings,
and to investigate the theory that hatches from the r.h.s. of eq.~(\ref{eq:Z}).

Our goal is to compute S-matrix elements between `in' and `out' primed
fields, that we have identified with the observed (asymptotic)
mesons; their classical equations (which are the Klein-Gordon equations
for free fields) stay the
same as those of the original `in' and `out' fields of $\cal L$;
this in particular entails that the Klein-Gordon operators occurring inside the
reduction formul\ae (see for example \cite{Roman}) stay unchanged.
However, because of the global $a^2$ now
factorizing the action in the r.h.s. of eq.~(\ref{eq:Z})\l
- each (bare) propagator gets a factor $1/a^2$;\l
- each gauge coupling $g'$ (recall that $g'$ is the gauge coupling in the
rescaled Lagrangian, see eq.~(\ref{eq:GAUGECPL})) gets a factor $a^2$;\l
such that the parameter `counting' the number of loops is now $\hbar c/a^2$.\l
The first and second points make each external leg get a factor $1/a^2$
because the propagators on the external legs of the full Green functions
are no longer canceled by the derivative operators mentioned above,
this by a factor $1/a^2$.

This results in the following facts:\l
- for the same number of asymptotic states, for example for the leptonic
decays of mesons and $\pi^0 \rar \gamma\gamma$, the diagram with one loop (the
latter) will get a factor $1/a^2$ with respect to the tree-diagram (the
former);\l
- when considering two processes occurring at tree-level, with
respectively three and four external legs,
as for example leptonic and semi-leptonic decays, the latter having one
more external leg than the former, its amplitude has one more $1/a^2$ factor.

Those considerations bear upon the relative normalizations of the different
processes of concern to us.

The question of the absolute normalization of the $S$-matrix elements is left.
The simplest way to get it is by looking at the low energy limit of a
physical process, and the one which naturally comes to the mind in a theory
of weak interactions is the four-leptons coupling.
It should reproduce the Fermi interaction at low
energies, now between `primed' leptons, considered as the physical ones.
One finds
\be
{\cal N}\;{(a^2)^2} {1\over a^2} {({1\over a^2})^4} {8g'^2 \over g'^2 v'^2}
          = {{\cal N}\over a^4}{8\over a^2 v'^2}
          = {{\cal N}\over a^4}{8\over v^2} = {{\cal N}\over a^4}G_F,
\ee
where, in the r.h.s., the $\cal N$ is the normalization factor sought for,
the second factor is due to the two coupling constants,
the third to the internal $W$ propagator, the fourth to the four external
legs. We see that $\cal N$ is required to be $a^4$.

Once we have fixed the global normalization of one process,
(which correctly reproduces experimental data), the normalization
of other processes is fixed by the discussion above.
They are also found to fit experiment:
this is the case for the leptonic decays of mesons, where we recover the
`naive' (and correct) result of section \ref{sec:LEPTONIC}, the decay of the
pion into two photons, and the semi-leptonic decays of mesons.
Indeed, for the latter, we find the coupling  ($p_\mu$ being the kaon
momentum and the $8$ factor coming from $M_W^2 = g^2 v^2/8 = g'^2v'^2/8$)
\be
(1/a^2)\;s_\theta \, p_\mu\,{8\over v'^2} = s_\theta \, p_\mu\,{8\over v^2},
\ee
corresponding to $f_+ + f_- = 1$, $f_+$ and $f_-$ being the two customary
form factors of $K_{\ell 3}$ decays.

Finally, the physical amplitudes are obtained by:\l
- computing with the above rules, that is, putting as required by
eq.~(\ref{eq:Z}), a factor $a^2$ with each coupling constant $g'$,
a factor $1/a^2$ with each internal propagator, and a factor $1/a^2$ with
each external leg;\l
- then, multiplying the result by  a global normalization factor ${\cal N}
=a^4$.\l

With the above rules of calculation, none of these processes requires the
introduction of another scale of interaction.

{\em Remark:} the global normalization factor $a^4$ we also introduce in the
reduction formul\ae.
Take the simplest  $S$-matrix element, that between two identical
asymptotic mesons, depicted in fig.~3:
\figskip
\hskip 5.5cm\epsffile{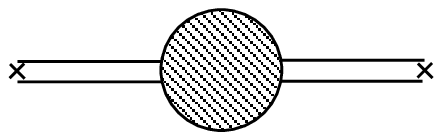}
\figskip
{\centerline{\em Fig.~3: `diagonal' S-matrix element.}}
\figskip
It has two external legs, giving, from the above considerations, two factors
$1/a^2$; so, the reduction {formul\ae}  bring it to:
(we have omitted the $\sqrt{Z}$'s normalization factors):
\be
_{out} \la \phi'(p) \vert \phi'(p) \ra _{in} =
 {{\cal N}\over a^4}\times
\ \mbox{`2-point\ proper-vertex'\ for\ interacting}\ \phi's.
\label{eq:SMAT}\ee
The two-point proper vertex is the bubble of fig.~3, where now no external
leg is there any longer. Perturbatively, the expansion of the above
$S$-matrix element begins, as `usual', by the inverse propagator, as we now
get it from eq.~(\ref{eq:Z}):
\be
{\cal N}{(p^2-m_\phi^2)^2\over a^2(p^2-m_\phi^2)} + \cdots =
a^2 (p^2 -m_\phi^2) + \cdots,
\ee
where, in the l.h.s., the two factors $(p^2- m_\phi^2)$ are the two
Klein-Gordon operators.

\subsection{Proposed identification of observed pseudoscalar
                 electroweak eigenstates.}\label{sec:PROPOSED}

The propositions below should be considered only as a first
and most simple attempt towards the determination of the physical states.
They are likely to  be refined in the future  by taking into account
more experimental data and constraints.

\subsubsection{The charged sector.}

By inspection of the decays evoked above, and anticipating the
results of the next section concerning non-leptonic decays into two pions,
we propose to identify the observed charged electroweak pseudoscalar
eigenstates as follows:
\bea
\pi^\pm &=& {v\over 4f \sqrt{2}}\left(
   c_\theta\,(\Phi^\pm + \Sigma^\pm) - s_\theta\,(i\,\Xi^\pm + \Omega^\pm)
                  \right),\nonumber\\
D_s^\pm &=& {v\over 4f \sqrt{2}}\left(
   c_\theta\,(\Phi^\pm - \Sigma^\pm) - s_\theta\,(i\,\Xi^\pm - \Omega^\pm)
                  \right),\nonumber\\
K^\pm  &=& {v\over 4f}(i\,c_\theta\,\Xi^\pm + s_\theta\,\Phi^\pm),
              \nonumber\\
D^\pm  &=& {v\over 4f}(c_\theta\,\Omega^\pm + s_\theta\,\Sigma^\pm).
\label{eq:MESONS}\eea
This ensures that
\bea
\pi^+ &=& {i\,v^2\over 2f \sqrt{2}N\mu^3}\ \bar u \gamma_5 d,\\
D_s^+ &=& {i\,v^2\over 2f \sqrt{2}N\mu^3}\ \bar c \gamma_5 s,\\
K^+   &=& {i\,v^2\over 4fN\mu^3}\ (\bar u \gamma_5 s - \bar c \gamma_5 d),\\
D^+   &=& {i\,v^2\over 4fN\mu^3}\ (\bar u \gamma_5 s + \bar c \gamma_5 d).
\label{eq:PI+}\eea
The relations (\ref{eq:MESONS}) above can be cast in the form
\be
\left(\ba{cc}
{1\over\sqrt{2}}(\pi^\pm + D_s^\pm) & {1\over\sqrt{2}}(\pi^\pm -D_s^\pm)\\
                                  K^\pm & D^\pm \ea\right)
= C^{-1}\times
{v\over 4f}\left(\ba{rr} \Phi^\pm & \Sigma^\pm \\
              i\,\Xi^\pm & \Omega^\pm \ea\right),
\label{eq:CABCH}
\ee
where $C$ is the Cabibbo mixing matrix (see eq.~(\ref{eq:CABIBBO})),
which is seen to act now at the mesonic level.

The above choice ensures that, (at least in the pseudoscalar sector that we
are investigating), the kinetic terms are
diagonal both in the basis of $\Phi, \Sigma, \Xi, \Omega$ and in that of the
mesons themselves (the corresponding mixing matrix is unitary).

\subsubsection{The neutral sector.}

We first exhibit the neutral pseudoscalar partners of the charged
eigenstates above, to which we add the pseudoscalar singlets in the `tilded'
representations, that we shall relate to the long-lived kaon
$K_L = K^0_1$ and to the $D^0_2$ meson; we specially study here  the case
of the  neutral $K$ mesons.
\bea
\pi^3 &=& {v\over 4f \sqrt{2}}\left(
   c_\theta\,(\Phi^3 + \Sigma^3) - s_\theta\,(i\,\Xi^3 + \Omega^3)
                  \right)\nonumber\\
    &=& i\,{v^2\over 4fN\mu^3}\left(
  c_\theta\,(\bar u \gamma_5 u - \bar d \gamma_5 d)
 - s_\theta\,(\bar u \gamma_5 c + \bar s \gamma_5 d) \right),\nonumber\\
d_s^3 &=& {v\over 4f \sqrt{2}}\left(
   c_\theta\,(\Phi^3 - \Sigma^3) - s_\theta\,(i\,\Xi^3 - \Omega^3) \right)
                                                                   \nonumber\\
  &=& i\,{v^2\over 4fN\mu^3}\left(
  c_\theta(\bar c \gamma_5 c - \bar s \gamma_5 s)
 + s_\theta(\bar c \gamma_5 u + \bar d \gamma_5 s) \right),\nonumber\\
k^0_1 &=& {v\over 4f}(i\,c_\theta\,\Xi^3 + s_\theta\,\Phi^3)
              \nonumber\\
   &=& i\,{v^2\over 4\sqrt{2}fN\mu^3}\left(
     c_\theta(\bar u \gamma_5 c - \bar c \gamma_5 u - \bar d \gamma_5 s
           + \bar s \gamma_5 d)
     + s_\theta( \bar u \gamma_5 u + \bar c \gamma_5 c - \bar d \gamma_5 d
           - \bar s \gamma_5 s)\right),\nonumber\\
d^0_2 &=& {v\over 4f}(c_\theta\,\Omega^3 + s_\theta\,\Sigma^3)\nonumber\\
   &=& i\,{v^2\over 4\sqrt{2}fN\mu^3}\left(
     c_\theta(\bar u \gamma_5 c + \bar c \gamma_5 u - \bar d \gamma_5 s
           - \bar s \gamma_5 d)
     + s_\theta( \bar u \gamma_5 u - \bar c \gamma_5 c + \bar d \gamma_5 d
           - \bar s \gamma_5 s)\right),\nonumber\\
& &
\label{eq:NEUTRAL}\eea
to which we add the `singlets' of the `tilded' representations that we
associate with $K^0_2$ and $D^0_1$:
\bea
k^0_2 &=& i\,{v\over 4f} (c_\theta\,\tilde\Omega^0 + s_\theta\,\tilde\Sigma^0)
              \nonumber\\
   &=& i\,{v^2\over 4\sqrt{2}fN\mu^3}\left(
     c_\theta(\bar u \gamma_5 c + \bar c \gamma_5 u + \bar d \gamma_5 s
         + \bar s \gamma_5 d)
    + s_\theta( \bar u \gamma_5 u - \bar c \gamma_5 c - \bar d \gamma_5 d
         + \bar s \gamma_5 s)\right),\nonumber\\
d^0_1 &=& {v\over 4f}(c_\theta\,\tilde\Xi^0 -i\, s_\theta\,\tilde\Phi^0)
                     \nonumber\\
      &=& i\,{v^2\over 4\sqrt{2}fN\mu^3}\left(
     c_\theta(\bar u \gamma_5 c - \bar c \gamma_5 u + \bar d \gamma_5 s
         - \bar s \gamma_5 d)
    - s_\theta(\bar u \gamma_5 u + \bar c \gamma_5 c + \bar d \gamma_5 d
         + \bar s \gamma_5 s)\right).\nonumber\\
& &
\label{eq:KSDL}\eea
The definition of $k^0_2$ is seen to introduce a non-diagonal mixing between
the kinetic terms of
$\tilde\Omega$ and $\tilde\Sigma$ which will have to be canceled by correctly
choosing the scalar eigenstates (we recall that the scalar mesons being
likely to undergo strong decays, we will not study them here).

We did not give the states in eqs.~(\ref{eq:NEUTRAL}, \ref{eq:KSDL})
above their usual names (using for example lower case letters) as they are
not exactly what are considered as  the states $\pi^0, K^0_1=K_L,
K^0_2=K_S, D^0_1, D^0_2$.

The first thing that can be noticed is that the neutral electroweak partner
of the charged pions, $\pi^3$, is not its own
antiparticle, nor is $d_s^3$, which belongs to the same representation as
$D_s^\pm$; also,  $(k^0_1 + k^0_2)$ and $(k^0_2 - k^0_1)$ are
not antiparticle of each other. We have:
\bea
{1\over\sqrt{2}}(k^0_1 + k^0_2) &=& i\,{v^2\over 4f N\mu^3}\left(
     c_\theta\,(\bar u \gamma_5 c + \bar s \gamma_5 d)
    +s_\theta\,(\bar u \gamma_5 u - \bar d \gamma_5 d)\right),  \nonumber\\
{1\over\sqrt{2}}(k^0_2 - k^0_1) &=& i\,{v^2\over 4fN\mu^3}\left(
     c_\theta\,(\bar c \gamma_5 u + \bar d \gamma_5 s)
    + s_\theta\,(\bar s \gamma_5 s - \bar c \gamma_5 c)\right).
\label{eq:KLKS}\eea
Instead, from eqs.~(\ref{eq:NEUTRAL}, \ref{eq:KLKS}, \ref{eq:KSDL}),
one has
\bea
& & \left({c_\theta\over\sqrt{2}}(k^0_1 + k^0_2) -s_\theta\,\pi^3\right)
    = i\,{v^2\over 4f N\mu^3}(\bar u \gamma_5 c + \bar s \gamma_5 d) = K^0
                                             \nonumber\\
&\mbox{and}& \nonumber\\
& &  \left({c_\theta\over\sqrt{2}}(k^0_2 - k^0_1) + s_\theta\,d_s^3\right)
  = i\,{v^2\over 4f N\mu^3}(\bar c \gamma_5 u + \bar d \gamma_5 s)=\bar K^0
\label{eq:ANTI}\eea
are antiparticle of each other; we thus naturally associate them with the
$K^0$ and $\bar K^0$ `strong' eigenstates (note that, as announced in
section \ref{sec:PRELIM}, they differ from their customary definitions in
terms of the $d$ and $s$ quarks only).
Also, of the two `orthogonal' states
\bea
& & \left(c_\theta\,\pi^3 + {s_\theta\over\sqrt{2}}(k^0_1 + k^0_2)\right)
  = i\,{v^2\over 4f N\mu^3}(\bar u \gamma_5 u - \bar d \gamma_5 d)= \pi^0
                                           \nonumber\\
&\mbox{and}& \nonumber\\
& & \left(c_\theta\,d_s^3 - {s_\theta\over\sqrt{2}}(k^0_2 - k^0_1)\right)
   =  i\,{v^2\over 4f N\mu^3}(\bar c \gamma_5 c - \bar s \gamma_5 s)= \chi^0,
\label{eq:ORTHO}\eea
the first corresponds  to the `strong' $\pi^0$, and the second should
be degenerate in mass with the $D_s^\pm$ mesons. $d_s^3$ is likely to be
the $\chi(1910)$ \cite{ParticleData} (see below).

The equations (\ref{eq:ANTI}, \ref{eq:ORTHO}) above can be cast in a form
similar to eq.~(\ref{eq:CABCH})
\be
\left( \ba{cc}  \pi^0 & \bar K^0 \\
                  K^0 & \chi^0    \ea\right) =
           C \times \left( \ba{cc} \pi^3 & {1\over\sqrt{2}}(k^0_2 - k^0_1) \\
                       {1\over\sqrt{2}}(k^0_1 + k^0_2) & d_s^3 \ea\right).
\label{eq:CABNEUT}\ee
So, while the kinetic terms
for the electroweak representations of section \ref{sec:REPS} are naturally
diagonal in the `strong' eigenstates (and we have been careful that it stays
the same after our change of basis above), the same thing does not occur for
the mass terms: the neutral electroweak partners of the charged (`strong' or
`weak') eigenstates  no longer have simple relations by charge conjugation,
and mixing terms appear.

The detailed study of the neutral sector will be the subject of a forthcoming
work. It will also deal with the other neutral states, those corresponding
to the $D$ mesons, just mentioned here, and with the mesons of the `$\eta$'
type, the mixing of which are known to be rather subtle.

Let us however give here a very simple example of the mechanism that
operates. It has furthermore the nice
property of yielding mass relations between charged and neutral (electroweak)
pions, and between the $D_s$ mesons and the $\chi(1910)$, both well
verified experimentally. We only use the charge independence of strong
interactions which forces all `strong' pions to have the same mass.
Such a mass term for the strong pions writes
\be
-{1\over 2} m_\pi^2 (2\pi^+\pi^- + {\pi^0}^2),
\ee
and can be rewritten, using eqs.~(\ref{eq:ANTI}, \ref{eq:ORTHO},
\ref{eq:CABNEUT}) above, and neglecting terms in $s_\theta^2$
\be
-{1\over 2} m_\pi^2 \left( 2\pi^+\pi^- + c_\theta^2\,\pi^3{\pi^3}^\dagger
     + c_\theta s_\theta\,(\pi^3 \bar K^0 + K^0{\pi^3}^\dagger) +\cdots\right).
\ee
This shows that the electroweak eigenstate $\pi^3$ behaves like a particle
with mass
\be
m_{\pi^3} = c_\theta\,m_{\pi^\pm},
\label{eq:MPI}
\ee
which has mixing with the neutral kaons; this last fact reflects the
already mentioned property that one cannot diagonalize the mass terms for
both strong and electroweak eigenstates. The relation (\ref{eq:MPI}) is
verified at better than $1$ percent.

In a similar way, one shows that
\be
m_{d_s^3} = c_\theta\,m_{D_s^\pm}.
\label{eq:MK}
\ee
If we suppose that $d_s^3$ is the $\chi(1910)$,
eq.~(\ref{eq:MK}) is verified at $0.4$ percent.

Those results are encouraging, and a more complete study is postponed to
a further work.

We conclude this section by a remark concerning the number of independent
mass scales, made simpler by working in the approximation where we neglect
$\sin\theta$ corrections. The reader has notices that $K_L$ and
$K_S$ are members of different representations, $\Xi$ for the former and
$\tilde\Omega$ for the latter, such that we could {\em a priori} expect two
different masses for theses mesons.
However, in this approximation $(K_L + K_S)$ and $(K_S - K_L)$ being
antiparticle of each other must have the same mass, meaning that the mass
scales associated to the two above representations of the electroweak gauge
group must be identical.
The same type of phenomenon occurs in the $D$ sector, such that the
{\em a priori} eight independent electroweak mass scales reduce to six.

{\em Remark:}  at the same approximation, it can immediately be
checked that the states $[K^+, (1/\sqrt{2}(K_L - K_S)]$ and
$[K^-, (1/\sqrt{2}(K_L+K_S)]$ are, as usual,
doublets of the strong isospin group defined by eq.~(\ref{eq:ISOSPIN}).
The same thing holds for $D$ mesons.

\section{{\boldmath $K \rar \pi\pi$} decays.}

We address here the question of the amplitudes for the
electroweak eigenstates $K_S = K^0_2, K^+$ (and $K_L = K^0_1$) to decay into
two  pions. The literature concerning those decays and the
so-called $\Delta I = 1/2$ rule is huge, and it is impossible to quote
fairly all of it. The reader will only find  a restricted choice of references,
putting the accent on the different techniques that have been used and on
the evolution of ideas and, inside them, many more.

In the decays under scrutiny, the final neutral pion is detected by
its photonic decays; so, in the following, one can safely calculate as if,
formally, it was a $(\bar u \gamma_5 u - \bar d \gamma_5 d)$ bound state;
indeed its other components have no projection on $\Phi^3$ and thus do not
decay into two photons (see the previous sections). This simplifies the
computations. Also,  as they are not `suppressed', the decays of $K_S$
will be computed by
taking this meson equal to the singlet of $\tilde\Omega$, neglecting
corrections in higher orders in $s_\theta$.

The three types of diagrams which give rise to such processes are described
in figs.~4, 5 and 6 below (the S and P symbols denote respectively scalar and
pseudoscalar states in all figures below).
\figskip
\vbox{
\hskip 4cm\epsffile{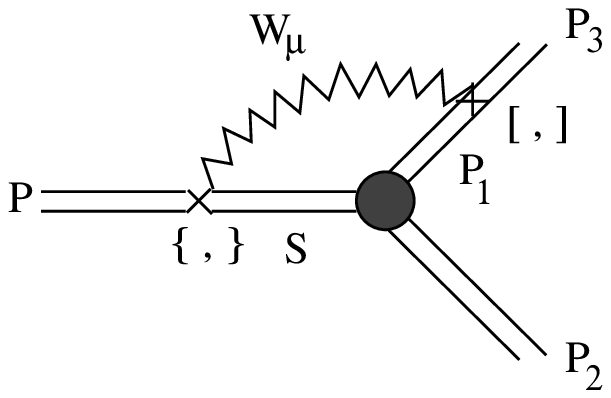}
\figskip
{\centerline{\em Fig.~4: first type of  contribution to $K \rar \pi \pi$.}}}
\figskip
\vbox{
\hskip 4cm\epsffile{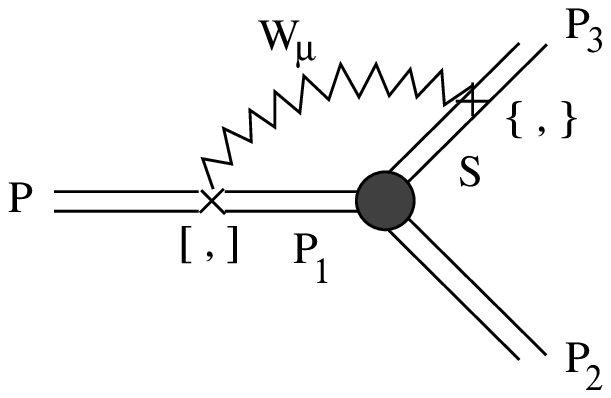}
\figskip
{\centerline{\em Fig.~5: second type of  contribution to $K \rar \pi \pi$.}}}
\figskip
\vbox{
\hskip 4cm\epsffile{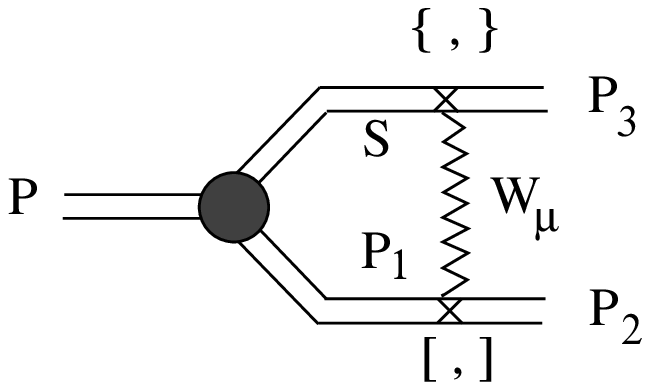}
\figskip
{\centerline{\em Fig.~6: third type of  contribution to $K \rar \pi \pi$.}}}
\figskip
The darkened bubble corresponds to the creation, by
strong interactions, (conserving parity), of
a $q\bar q$ quark pair, diagonal in flavour, transforming a pseudoscalar
bound state into one scalar plus one pseudoscalar, and a scalar bound state
into two pseudoscalars (the transformation of one scalar into two scalars is
of course possible but does not correspond here to a process of interest).
Each cross stands for a weak vertex, connecting either a gauge field and
two pseudoscalars or a gauge field, one scalar and one pseudoscalar; those
vertices arise in the kinetic terms for the mesons: in the first case, the
gauge generator acts on the composite state by commutation, transforming a
pseudoscalar into another pseudoscalar, and in the second case it acts by
anticommutation, transforming a pseudoscalar into a scalar, or the opposite
(see eqs.~(\ref{eq:GA1},\ref{eq:GA2})).
The mode of action of the gauge group has also been symbolized on the three
figures. The contributions where the $W$ propagator is attached to the same
ingoing or outgoing particle identically vanish due to the identity
symbolically described in fig.~7 and which can be checked easily.
\figskip
\vbox{
\hskip 2cm\epsffile{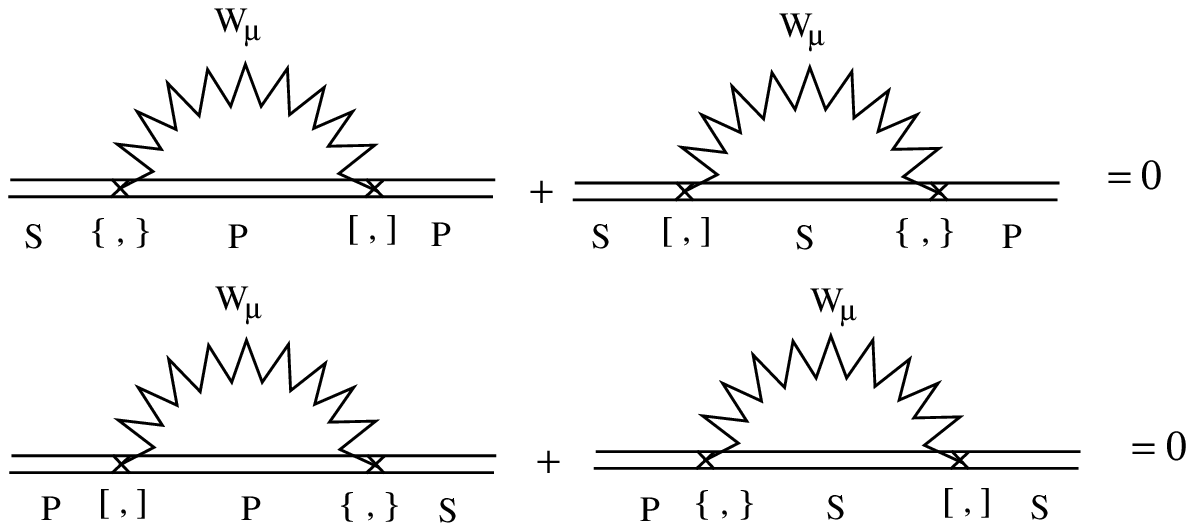}
\vskip 0cm
{\centerline{\em Fig.~7: identity making the fourth type of contribution to
                          $K\rar \pi\pi$ vanish.}}}
\figskip
The diagrams of the type of fig.~8 could {\em a priori} appear since the
kaons, as defined in eqs.~(\ref{eq:MESONS}), have a non-vanishing component
on $\Phi$, which allows a $K-W_\mu$
transition (see section \ref{sec:LEPTONIC}). However the coupling of the $W$
to two mesons happens to be antisymmetric in the mesons, which makes it
vanish in this case since the two mesons are pions which must, at the
opposite, be symmetrized because of Bose statistics.
\figskip
\hskip 4cm\epsffile{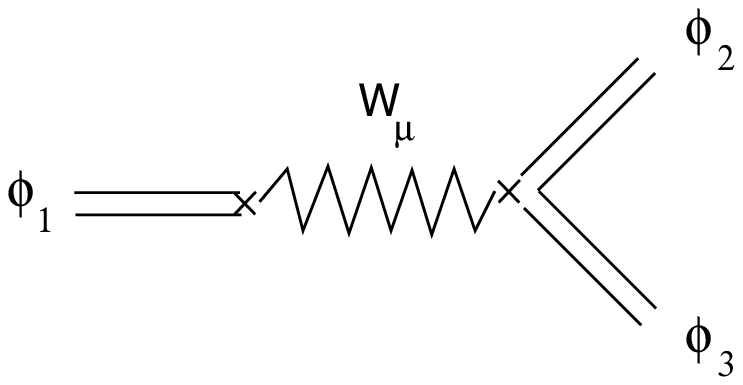}
\figskip
{\centerline{\em Fig.~8: diagrams that do not occur in
                                                   $K \rar \pi \pi$ decays.}}
\figskip
Because of the unknown value of the `strong' vertex, we cannot give precise
numbers, but will only unravel the suppression of certain types of decays,
from a new point of view, going in particular beyond
any `picture' like the `quark-spectator model', or any `factorization'-like
hypothesis. We indeed use composite representations of the
electroweak group and do not have to raise the question of which
line of quark the weak boson is attached to.

Let us work in the approximation where we
neglect all effects of flavour symmetry breaking, which means that:\l
- all strong vertices will be considered to have the same value;\l
- all intermediate states will be considered to propagate in the same
way (we neglect the mass splittings).\l
The dynamics is consequently maximally simplified, and becomes the same
for all diagrams in figs.~4, 5, 6 with the same ingoing and outgoing states,
up to a `$-$' sign between fig.~5
and fig.~6: this can be easily seen if one remembers that the weak vertices
involve a derivative of the incoming meson, and that
fig.~5 and fig.~6 can be deduced from one another by exchanging the incoming
pseudoscalar $P$ and the final $P2$, together with changing the
sign of their momenta: one being attached to a `strong' vertex, this change is
considered {\em a priori} not to alter it, while the other, attached to a weak
vertex, will change the sign of this vertex.
Our argumentation thus relies on pure group theory, calculating
commutators and anticommutators between electroweak generators and matrices
determining the scalar bound states, and on
the hypothesis concerning strong interactions: they act by flavour-diagonal
quark pair creation (thus conserving isospin, strangeness etc \ldots), and
are flavour independent (the same for all scalar and pseudoscalar bound
states).

We study the decays of $\Xi^+$, $\Xi^3$ and $\tilde\Omega^0$, corresponding
to $K^+$, $K_L$ and $K_S$  electroweak eigenstates into two pions;
that is, we suppose that the strongly created initial quark pair has been
collimated to select the particles with the kaon mass.
The results are the following:\l
$\ast$ for $K^+ \rar \pi^+ \pi^0$:\l
\quad - for the $\Xi$ component of $K^+$:  the diagrams of figs.~5 and 6
contribute,
the gauge field being $W_\mu^3$, and exactly cancel in the limit of exact
$SU(4)$ symmetry (there is no contribution with $W_\mu^\pm$);\l
\quad - for the $\Phi$ component of $K^+$: the diagrams of fig.~4 do not
contribute; the diagrams of figs.~5 and 6 again cancel, the gauge fields being
$W_\mu^3$ and $W_\mu^\pm$;\l
$\ast$ for $K_S \rar \pi^+\pi^-$, only the diagram of fig.~4 now contributes,
with gauge fields $W_\mu^\pm$ (there is no contribution with $W_\mu^3$);\l
$\ast$  for $K_L \rar \pi^+\pi^-$,  cancelations between figs.~5 and 6 again
occur, like for $K^+ \rar \pi^+\pi^0$, with an inversion of the roles of
$W_\mu^3$ and $W_\mu^\pm$.

All decays depend on the Cabibbo angle like $\cos\theta_c \sin\theta_c$.

We thus conclude that, as observed experimentally, and unlike
what is suggested by naive `quark spectator' models, $K^\pm \rar \pi^\pm
\pi^0$ and $K_L \rar \pi^+ \pi^-$ are strongly suppressed with respect to
$K_S \rar \pi^+\pi^-$. The non-complete vanishing of the first two should be
attributed to the breaking of the $SU(4)$ symmetry. Dynamical inputs
concerning strong interactions would be necessary to go beyond this
qualitative argumentation, but we think it to be illusory in our present
state of knowledge, since the would-be candidate for a theory of strong
interactions cannot yet tell anything about that of observed particles (see
for example the introduction of \cite{Feynman}).

The only dynamical remark which can be suggested is that the convergence of
the above diagrams requires the strong vertex to have a behaviour in inverse
power of the incoming meson momentum. Considering this as a hint that strong
interactions do indeed become stronger at low energy is probably, however,
going much too fast. It can anyhow be kept in mind for further studies.

\section{Conclusion.}

In continuation of the program started in
refs.~\cite{Machet1,Machet2,BellonMachet},
this work was devoted to the study of composite representations of the
electroweak symmetry group, in the case of two
generations of quarks and leptons. It shows that modifying the
conventional beliefs concerning the quark content of pseudoscalar mesons
provides a new insight into
the physics of those particles. In particular, the decays of kaons into two
pions are clarified. The possibility of radically departing from a pure
quark picture  by writing a gauge
theory for scalar fields which are composite representations of the symmetry
group is specially attractive. This procedure can be extended to the case
of three generations. Of course, the enormous (and ever growing) amount
of experimental data concerning heavy mesons makes the determination of the
observed states as electroweak eigenstates much more difficult,
and the choice should be first guided by principal facts and intuition.

Difficulties arise when dealing with
processes where not only electroweak, but also strong interactions, are
concerned, like the non-leptonic decays of pseudoscalar mesons. The need for
a real theory of strong interactions is pressing, to enable dynamical
computations. They also require the knowledge of the spectrum of particles
running in the internal lines of figs.~4, 5, 6, including elusive scalars,
to be able to go beyond the approximation of exact $SU(N)$ symmetry.
A departure from this symmetry for the quark condensates can also
undoubtedly enrich the phenomenology of the model. But one has always to
keep in mind that the goal is not to increase the number of arbitrary
parameters, but rather to fully exploit the few `unavoidable' ones, like
the mixing angles. We hope to have made here a step in that direction,
since, in particular, the phenomenological `quark mass' parameters no longer
appear.

The picture that springs out of this approach is that of the preeminence of
the chiral group of strong interactions, broken down to the electroweak
group which, in particular, determines the mass spectrum of observed
electroweak eigenstates. The latter is itself spontaneously broken down to
electromagnetism and, inside each electroweak multiplet, a fine mass
structure will result from pure electromagnetic interactions. A disturbing
question is that only a subgroup of the chiral group seems, up to now, to be
a local symmetry. We have already evoked in the core of the paper the need
for  gauging  the whole group: this would allow a more symmetric
treatment of the constraints introduced in the Feynman path integral to
express the compositeness of the  mesons, the ideal situation being that
every observed pseudoscalar or scalar  meson  be the third polarization
of a massive gauge field. However, the mass hierarchies physically
observed undoubtedly makes this a challenge, in need of a
mechanism yet to be uncovered.

Coming back, finally, to more practical concerns,
it must have been obvious to the reader that we have always deliberately
ignored all `quark diagrams' which were, up to now,
 the only ones  looked at. We have thus supposed that ours have `replaced'
the others, meaning that the quark degrees of freedom have been frozen and
transmuted into those of the mesons. A more formal argumentation in favour of
this point of view can be found in refs.~\cite{Machet1,Machet2}, where it is
shown that quantizing a theory with composite, non-independent,
degrees of freedom (remember that in the Glashow-Weinberg-Salam
model, one integrates both on the quarks and on the scalars, and that now
the latter are made of the former), requires adding constraints in the
Feynman path integral, which bring back to the correct
number of degrees of freedom: these constraints can be exponentiated
into an additional effective Lagrangian which freezes the quark
degrees of freedom by giving them infinite masses. Those consequently
decouple, with a subtlety linked to one-loop processes, the anomaly and
the photonic decay of the pion, evoked in section \ref{sec:ANOM};
the latter is recovered from the constraints
themselves, though the theory is now anomaly-free. A similar decoupling
occurs for the Higgs boson, which we predict to be unobservable.
Formal problems are still to be investigated and will be the subject of
forthcoming works.

\bigskip
\begin{em}
\underline{Acknowledgments}:
Many thanks are due to my colleagues in LPTHE, specially to J.~Avan and
M.~Talon  for their help to improve the manuscript, M.~Bellon for providing
us with a very efficient computer system, and C. M.~Viallet for his assistance
with the Maple programming language.
\end{em}
\newpage\null
\listoffigures
\bigskip
\begin{em}
Fig.~1:  Leptonic decay of a pseudoscalar meson;\l
Fig.~2:  Semi-leptonic decay of a meson;\l
Fig.~3:  `Diagonal' S-matrix element;\l
Fig.~4:  First type of contribution to $K \rar \pi\pi$;\l
Fig.~5:  Second type of contribution to $K \rar \pi\pi$;\l
Fig.~6:  Third type of contribution to $K \rar \pi\pi$;\l
Fig.~7:  Identity making the fourth type of contribution to $K\rar\pi\pi$
                                                                     vanish;\l
Fig.~8:  Diagrams that do not occur in $K \rar \pi \pi$ decays.
\end{em}

\newpage\null
\begin{em}

\end{em}

\end{document}